\documentclass[aps,prl,twocolumn,superscriptaddress]{revtex4}
\usepackage{amssymb}
\usepackage{graphicx}
\usepackage{dcolumn}
\usepackage{bm}
\usepackage{siunitx}

\begin{document}

\preprint{APS/123-QED}

\title{High-resolution electric field imaging \\ based on intermittent-contact mode scanning NV center electrometry}
\author{Zhi Cheng}
\altaffiliation{These authors contributed equally to this work.}
\affiliation{School of Physical Sciences, University of Science and Technology of China, Hefei 230026, China}
\affiliation{Anhui Province Key Laboratory of Scientific Instrument Development and Application, University of Science and Technology of China, Hefei 230026, China}

\author{Zhiwei Yu}
\altaffiliation{These authors contributed equally to this work.}
\affiliation{School of Physical Sciences, University of Science and Technology of China, Hefei 230026, China}
\affiliation{Anhui Province Key Laboratory of Scientific Instrument Development and Application, University of Science and Technology of China, Hefei 230026, China}

\author{Mengqi Wang}
\affiliation{School of Physical Sciences, University of Science and Technology of China, Hefei 230026, China}
\affiliation{Anhui Province Key Laboratory of Scientific Instrument Development and Application, University of Science and Technology of China, Hefei 230026, China}

\author{Lingfeng Yang}
\affiliation{School of Physical Sciences, University of Science and Technology of China, Hefei 230026, China}
\affiliation{Anhui Province Key Laboratory of Scientific Instrument Development and Application, University of Science and Technology of China, Hefei 230026, China}

\author{Zihao Cui}
\affiliation{School of Physical Sciences, University of Science and Technology of China, Hefei 230026, China}
\affiliation{Anhui Province Key Laboratory of Scientific Instrument Development and Application, University of Science and Technology of China, Hefei 230026, China}

\author{Ya Wang}
\affiliation{School of Physical Sciences, University of Science and Technology of China, Hefei 230026, China}
\affiliation{Anhui Province Key Laboratory of Scientific Instrument Development and Application, University of Science and Technology of China, Hefei 230026, China}
\affiliation{Hefei National Laboratory, University of Science and Technology of China, Hefei 230088, China}

\author{Pengfei Wang}
\email{wpf@ustc.edu.cn}
\affiliation{School of Physical Sciences, University of Science and Technology of China, Hefei 230026, China}
\affiliation{Anhui Province Key Laboratory of Scientific Instrument Development and Application, University of Science and Technology of China, Hefei 230026, China}
\affiliation{Hefei National Laboratory, University of Science and Technology of China, Hefei 230088, China}

\date{\today}

\begin{abstract}
Scanning nitrogen-vacancy (NV) center electrometry has shown potential for quantitative quantum imaging of electric fields at the nanoscale. 
However, achieving nanoscale spatial resolution remains a challenge since employing gradiometry to overcome electrostatic screening causes resolution-limiting trade-offs including the averaging effect and the sensor-sample proximity.
Here, we demonstrate a scanning NV center protocol that achieves an enhanced spatial resolution of approximately \SI{10}{\nm}. 
We develop an axially symmetric probe with a sub-nanometer oscillating amplitude, which simultaneously provides robust intermittent-contact mode feedback and ensures close engagement between the diamond tip and the sample. As an example, we experimentally demonstrate a \SI{10}{\nm} spatial resolution on ferroelectric lithium niobate. 
Scanning NV center electrometry with this resolution can directly resolve the nanoscale polar textures and dynamics of emerging ferroelectrics, which commonly arise on the scale of tens of nanometers. 
\end{abstract}

\maketitle


The behavior of electric fields profoundly reflects material properties and device performance across a wide range of disciplines. 
Originating from intricate polarization or charge configurations, electric fields are central to various phenomena and applications such as interfacial ferroelectricity\cite{fei_ferroelectric_2018, cheema_enhanced_2020, rogee_ferroelectricity_2022}, topological structures\cite{das_observation_2019,xue_observation_2025}, multiferroic coupling\cite{seki_observation_2012,chaudron_electric-field-induced_2024,masuda_electric_2021}, and electronic device degradation\cite{huang_direct_2021}.
Therefore, visualizing nanoscale electric fields will deepen the comprehension of polar texture for topological structures\cite{nataf_domain-wall_2020,li_unusual_2025} and facilitate the optimization of semiconductor devices performance\cite{cao_electric_2021}.

In recent years, nitrogen-vacancy (NV) centers in diamond have been developed as quantum electrometers, offering exceptional sensitivity and spatial resolution under ambient conditions\cite{dolde_electric-field_2011, dolde_nanoscale_2014, mittiga_imaging_2018, li_nanoscale_2020}. 
These unique features make NV centers ideal probes for sensing electric fields across a broad range of applications. 
Pioneering works \cite{huxter_imaging_2023, qiu_nanoscale_2022} have shown the capability of scanning NV center electrometry for imaging stray electric fields from ferroelectric domains and micro-fabricated electrodes, demonstrating a spatial resolution of approximately \SI{100}{\nm}. 
Despite these advances, exploring emerging ferroelectric phenomena unfold on the scale of tens of nanometers, such as the behavior of Van der Waals ferroelectrics\cite{vizner_stern_interfacial_2021,yasuda_stacking-engineered_2021}, remains a challenge for the spatial resolution of scanning NV center electrometry, as the mandatory use of gradiometry\cite{huxter_imaging_2023, qiu_nanoscale_2022, huxter_scanning_2022} to overcome electrostatic screening fundamentally couples the resolution to a distinct set of issues not present in direct sensing.

In this work, we analyze the factors that limit the spatial resolution of gradiometry: the averaging effect from the tip's mechanical oscillation in gradiometry, and the distance between the NV center and the sample. To address these issues, we present a scanning NV center electrometry that adopts an intermittent-contact mode. Our protocol not only dramatically reduces the mechanical oscillation amplitude from tens of nanometers to sub-nanometer scale but also permits closer probe-sample engagement due to stable feedback. We experimentally benchmark this method on a lithium niobate ferroelectric single crystal, achieving a spatial resolution of \SI{10}{\nm}.

\begin{figure}
\includegraphics{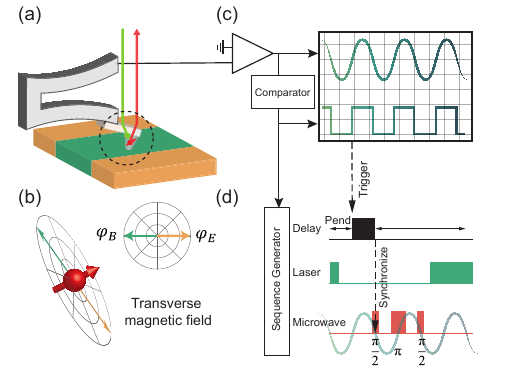}
\caption{\label{fig:setup}  Schematic of the experimental setup. (a) The scanning NV center microscope, which operates in the intermittent-contact mode. (b) NV center coordinate. The orientation $\varphi_B$ of the transverse magnetic field $B_\bot$ selects the projection of the electric field $\varphi_E$. (c) Conversion of the analog sinusoidal wave from the oscillating tuning fork into a digital square wave for triggering. (d) Implementation of the time synchronization between the spin-echo sequence and the probe's mechanical oscillation.}
\end{figure}

\begin{figure*}
\includegraphics{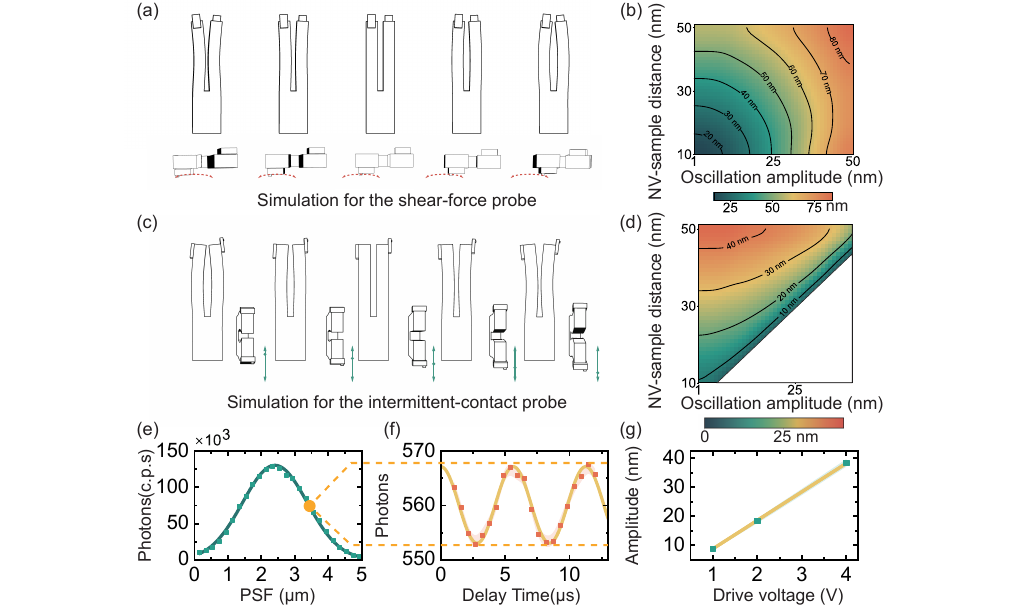}
\caption{\label{fig:probe}  (a),(c) Simulated tip trajectories for a probe oscillating at its second-order resonance frequency. (a) The probe in shear-force mode, whose primary oscillation axis is parallel to the sample, exhibits off-axis motion. (c) In contrast, the probe in intermittent-contact mode, whose primary oscillation axis is perpendicular to the sample, maintains linear harmonic motion. (b) Dependence of spatial resolution on the NV-sample distance and oscillation amplitude for gradiometry in shear-force mode. (d) Dependence of spatial resolution on the NV-sample distance and oscillation amplitude for gradiometry in intermittent-contact mode. The blank area indicates where the oscillation amplitude exceeds the NV-sample distance. (e) Photoluminescence distribution of the NV center along the oscillation axis, captured by confocal microscopy. The calibration was performed at the position of maximum gradient. (f) Time-resolved photoluminescence from the NV center, which oscillates sinusoidally at the same frequency as the tuning fork. (g) The calibrated oscillation amplitude shows a linear dependence on the driving voltage.
}
\end{figure*}
The NV center's Hamiltonian of the ground-state spin triplet under a bias magnetic field is given by \cite{doherty_nitrogen-vacancy_2013}:

\begin{eqnarray}
&&H_{gs} = D_{gs}S_{z}^{2} + \gamma_{e}B_{\bot}\left( {{\cos\varphi_{B}}S_{x} + {\sin\varphi_{B}}S_{y}} \right) \nonumber\\
&&+ d_{\bot}E_{\bot}\left\lbrack {{\cos\varphi_{E}}\left( {S_{x}^{2} - S_{y}^{2}} \right) - {\sin\varphi_{E}}\left( {S_{x}S_{y} + S_{y}S_{x}} \right)} \right\rbrack
\end{eqnarray}
where $D_{\rm{gs}}$ is the zero-field splitting, $\gamma_{\rm{e}}$ is the gyromagnetic ratio of electron spin. $B_{\bot} \cos\varphi_{B}$, $B_{\bot} \sin\varphi_{B}$ are the $x$ and $y$ components of the bias magnetic field which is transverse to the NV axis. $\mathbf{S}=\left(S_{x},\ S_{y},\ S_{z}\right)$ are the electron spin-1 operators. $d_\bot=(17\pm3)\ \rm{Hz\ cm\ V^{-1}}$ is the NV centers' non-axial electric dipole moment. $E_{x_{NV}} = E_{\bot} \cos\varphi_{E}$, $E_{y_{NV}} = E_{\bot} \sin\varphi_{E}$ are the electric field's $x$ and $y$ components. Due to the Stark effect, the spin transition frequencies experience a shift:
\begin{equation}
\Delta\omega_{\pm} = \mp 2\pi d_{\bot}E_{\bot}{\cos\left( {2\varphi_{B} + \varphi_{E}} \right)}
\end{equation}

To perform electrometry using shallow NV centers, the gradiometry scheme is employed to circumvent the electrostatic screening effect from the diamond surface(Fig.~\ref{fig:setup}): the oscillation of the tip converts the non-uniform DC electric field in the lab frame into an AC field in the tip's frame, and the amplitude of the AC field can be measured by dynamical decoupling protocols. 

In previous works\cite{huxter_imaging_2023, qiu_nanoscale_2022}, the probe designed for gradiometry works in shear-force mode, and oscillates at the second harmonic frequency to surpass the critical screening threshold ($ \sim $\SI{100}{\kHz}). According to our simulation, the shear-force mode probe\cite{guo_flexible_2021} suffers from structural asymmetries that give rise to a lateral oscillation component perpendicular to the primary axis at the second harmonic frequency(Fig.~\ref{fig:probe}(a)), even in absence of sample contact. This non-linear behavior compromises measurement accuracy in gradiometry and limits the minimum probe-sample distance, thereby degrading spatial resolution. Therefore, we adopt the intermittent-mode geometry for gradiometry because of the axial symmetry allows for perfect linearity during oscillation (Fig.~\ref{fig:probe}(c)).

We analyze two factors determining the spatial resolution for gradiometry in intermittent-contact mode: the averaging effect due to the oscillation of the tips and the impact of the distance between the NV center and the sample. As illustated in Fig.~\ref{fig:probe}(d), the resolution\cite{supp} significantly improves as the NV center approaches the sample. However, in the intermittent-contact mode, the NV-sample distance and the probe's oscillation amplitude are entirely dependent: the NV-sample distance must be greater than the oscillation amplitude. Therefore, to reduce the probe-sample distance, we decrease the probe's oscillation amplitude to achieve high spatial resolution. 

By careful design of our probe\cite{supp}, the oscillation amplitude is estimated to be smaller than \SI{1}{\nm}, which is nearly two orders below the optical diffraction limit, making precise measurement of the oscillation amplitude a challenge. We employ a time-resolved optical method, which consists of the following steps:
(1) Confocal scanning imaging along the oscillation direction of the NV center, yielding the distribution of NV fluorescence (Fig.~\ref{fig:probe}(e)):
\begin{equation}
h(z) = s_0 + \frac{s}{w \sqrt{\pi/2}} e^{-2((z-z_0)/w)^2}
\end{equation}
where \( s_0 \) is the background photon count, and \( s \) and \( w \) characterize the intensity and broadening.
(2) Focus the objective on the half-maximum position of the NV fluorescence spot, which ensures that the gradient of the fluorescence distribution is maximized (Fig.~\ref{fig:probe}(e)).
(3) Driving the NV center to oscillate:
\begin{equation}
z(t) = A \sin(\omega t + \phi) + z_0
\end{equation}
where \( A \) is the oscillation amplitude to be measured, \( \omega \) is the oscillation frequency, \( \phi \) is the phase, and \( z_0 \) is the averaging position in the laboratory frame.
Given that the oscillation amplitude is approximately two orders of magnitude smaller than the broadening of the fluorescence distribution ($\sim$ \SI{2.5}{\um}), the optical intensity can be considered linearly dependent on the oscillation position.
The linear approximation of the fluorescence distribution at the half-maximum width is:
\begin{equation}
h(z) = k(z - z_0) + s_0
\end{equation}
Consequently, the NV center’s harmonic motion manifests as a sinusoidal function in the time domain (Fig.~\ref{fig:probe}(f)). Using the fitted maximum and minimum photons during oscillation, we derive the following equation:
\begin{equation}
\frac{h_{\max}}{h_{\min}} = \frac{kz_{\max} + s_0}{kz_{\min} + s_0} \rightarrow \frac{h_{\max}}{h_{\min}} = \frac{k(z_0 + A) + s_0}{k(z_0 - A) + s_0}
\end{equation}
Here, the identities \( z_{\max} \to z_0 + A \) and \( z_{\min} \to z_0 - A \) hold due to the monotonicity of \( h(z) = k(z - z_0) + s_0 \). Since \( A \) is the only variable, solving this equation yields the oscillation amplitude of the probe.

We experimentally measure the mechanical oscillation amplitudes of the probe at different driving voltages(Fig.~\ref{fig:probe}(g)). The results exhibit an excellent linear dependence between the driving voltage and the amplitude. Therefore, in this voltage range, the oscillation amplitude can be directly determined using the established linear relationship. The oscillation amplitude is fixed at \SI{0.8}{\nm} in following experiments, corresponding to a driving voltage of \SI{100}{\mV}.

\begin{figure}
\includegraphics{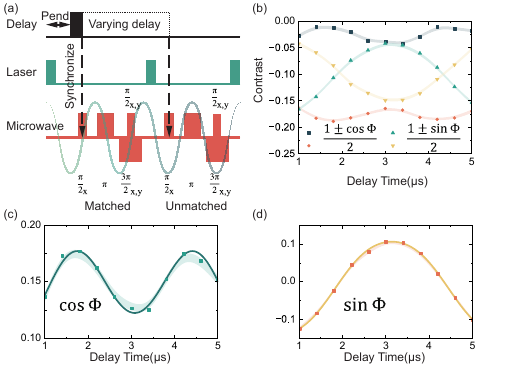}
\caption{\label{fig:sequence} The spin-echo protocol for electrometry. (a) The synchronized spin-echo sequence for gradiometry. The spin-echo sequence is triggered by the probe's mechanical oscillation after a delay $\tau_w$. (b) Experimental validation of the synchronized spin-echo sequence by sweeping the delay $\tau_w$. (c),(d)Respective fits of the experimental data using $\cos \Phi$ and $\sin \Phi$.}
\end{figure}

In our experiment, the amplitude of this AC field is measured by applying a spin-echo sequence whose total echo time equals to the oscillation period(Fig.~\ref{fig:sequence}). This protocol maps the AC field amplitude to an accumulated quantum phase. To maximize the sensitivity of gradiometry, the $\pi$ pulses of the spin-echo sequence must be synchronized to the zero-crossings of the AC field. To compensate for an unknown electronic delay $\tau_e$ due to the electronic circuit, we experimentally perform phase calibration by inserting a waiting time $\tau_w$ after the trigger from the comparator and sweeping its value to find the optimal timing.

\begin{figure*}
\includegraphics{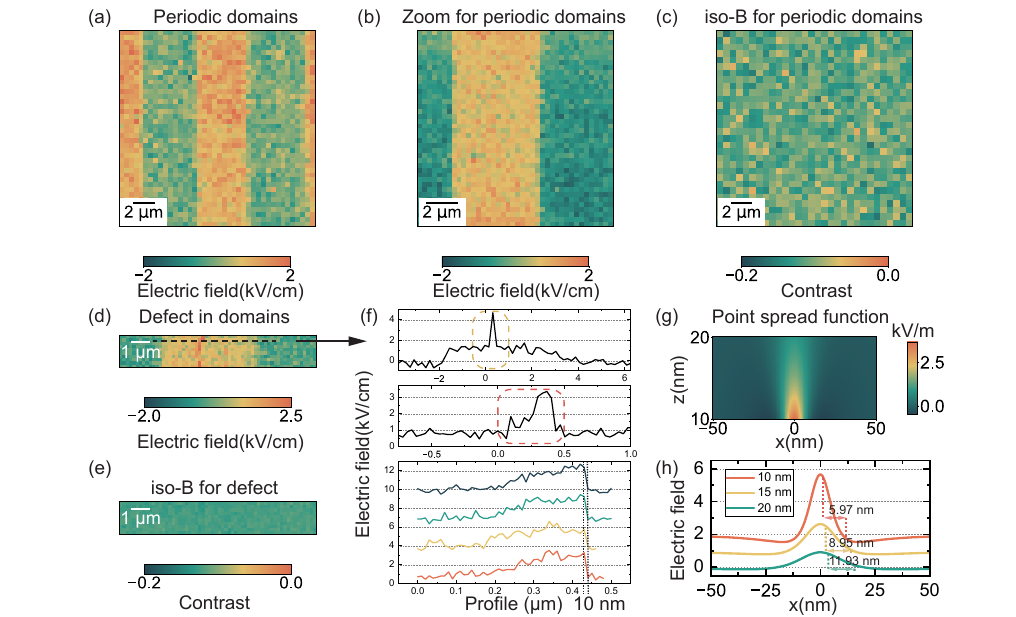}
\caption{\label{fig:exp} Electric field imaging of ferroelectric domains in lithium niobate. (a) A electric field map of a periodically-poled lithium niobate sample, acquired over an \SI{18}{\um} area. (b) A zoom-in scan for the detailed structure of the domain walls. (c) A pulsed ODMR measurement of the same area shown in (a). (d) A localized defect in a ferroelectric domain, used to benchmark the spatial resolution. (e) Pulsed ODMR map of the sample area in (d), confirming the electric-field nature of the feature. (f) Line profiles taken across the defect. The profiles are taken from scanning experiments ranging \SI{10}{\um}, \SI{2}{\um}, and \SI{0.5}{\um}; Complete profiles are shown in the supplementary material\cite{supp}. The 10\%-90\% edge-width criterion yields a spatial resolution of $\sim$ \SI{10}{\nm}.  (g) The point spread function for gradiometry of a $\delta(x)$ charge distribution. (h) The line profile for the point spread function. }
\end{figure*}

The accumulated quantum phase versus the total time delay $\tau_t=\tau_e+\tau_w$ is:
\begin{equation}
\Phi(\tau_t) = \int_{-\tau/2}^{\tau/2} \Theta(t) \cdot 2\pi d_\perp E_{\rm{AC}} \sin(\omega t + \omega \tau_t)\,dt\label{eq:quantum_phase}
\end{equation}
where $\omega$ is the oscillation frequency, $f=\omega/2\pi$, $\Theta(t)$ is the sign function, $\tau$ is the free evolution time for spin-echo sequence. $E_{\rm{AC}}$ is the amplitude of the AC electric field on NV center. $E_{\rm{AC}}$ is kept as a constant during the calibration of delay $\tau_w$.

Integration.~(\ref{eq:quantum_phase}) can be calculated as:
\begin{equation}
\Phi(\tau_t) = 4d_\perp E_{\rm{AC}} \frac{\sin^2(\pi f \tau/2)}{f} \cos(\omega \tau_t)
\end{equation}
This expression shows that $\tau_t$ modulates the accumulated quantum phase $\Phi$, and the maximum sensitivity is obtained when $\tau_t = \frac{n\pi}{\omega}$, where $n \in \mathbb{N}$. Since the $\tau_e$ is a constant which is determined by the electronics, we can adjust $\tau_w$ to make $\tau_e + \tau_w = \frac{n\pi}{\omega}$ to optimize the sensitivity.

We perform differential measurements using $\pi/2_y$ and $3 \pi/2_y$ to project $\Phi$ to population probabilities:
\begin{equation}
P_{s+} = \frac{1}{2}\left(1 + \sin \Phi \right)\label{eq:sine_result_pos}
\end{equation}
\begin{equation}
P_{s-} = \frac{1}{2}\left(1 - \sin \Phi \right)\label{eq:sine_result_neg}
\end{equation}

Subtracting these yields:
\begin{eqnarray}
P_s = \sin \Phi = \sin\left[4 d_\perp E_{\rm{AC}} \frac{\sin^2(\pi f \tau/2)}{f} \cos(\omega \tau_t) \right]\label{eq:sine_result}
\end{eqnarray}

Similarly, if using $\pi/2_x$ and $3 \pi/2_x$ to project $\Phi$, the population probabilities are:
\begin{equation}
P_{c+} = \frac{1}{2}\left(1 + \cos \Phi \right)\label{eq:cos_result_pos},
\end{equation}
\begin{equation}
P_{c-} = \frac{1}{2}\left(1 - \cos \Phi \right)\label{eq:cos_result_neg},
\end{equation}

\begin{eqnarray}
P_c = \cos \Phi = \cos\left[4 d_\perp E_{\rm{AC}} \frac{\sin^2(\pi f \tau/2)}{f} \cos(\omega \tau_t) \right]\label{eq:cos_result}
\end{eqnarray}

To validate our electrometry protocol, we performed experiments with a static electric field generated by nearby electrodes\cite{cheng_radio-frequency_2023}. As shown in (Fig.~\ref{fig:sequence}(b)), the measured data (green, yellow, red, and dark points), corresponding to final pulses of $\pi/2_y$, $3 \pi/2_y$, $\pi/2_x$ and $3 \pi/2_x$, perfectly trace the behaviors predicted by Eq.~(\ref{eq:sine_result_pos}), Eq.~(\ref{eq:sine_result_neg}), Eq.~(\ref{eq:cos_result_pos}), and Eq.~(\ref{eq:cos_result_neg}), respectively. Fitting the experimental data with Eq.~(\ref{eq:sine_result}) and Eq.~(\ref{eq:cos_result}) yields effective electric fields $E_{\rm{AC}}$ of \SI{2.36+-0.05}{\kV\per\cm} and \SI{2.61+-0.35}{\kV\per\cm}, respectively.

To demonstrate the imaging capabilities and benchmark the spatial resolution, we use a standard lithium niobate sample (PFM03, from NT-MDT Spectrum Instruments). This sample features a \SI{10}{\um} periodical striped pattern of antiparallel ferroelectric domains, with the spontaneous polarization oriented perpendicular to the surface.

An electric field gradient map is acquired over an \SI{18}{\um} $\times$ \SI{18}{\um} area. The result (Fig.~\ref{fig:exp}(a)) resolves a series of periodic stripes. A profile analysis of this pattern yields a periodicity of \SI{10}{\um}, which is consistent with that specified by piezoresponse force microscopy (PFM). We notice that the abrupt change of the electric field at a domain wall should produce a sharp transition for gradiometry. The absence of such distinct transition in our results (Fig.~\ref{fig:exp}(b)) might be attributed to the surface screening of the sample\cite{kalinin_surface-screening_2018} or physisorption on ferroelectrics\cite{li_direct_2008}, which likely smooth the electric field profile from a sharp step into a gradual transition. The surface screening on ferroelectric surfaces remains a challenge, even though our gradiometry can overcome the electrostatic screening originating from the diamond surface.

We acquire an electric field gradient map over an \SI{18}{\um} × \SI{18}{\um} area, which reveals a series of periodic stripes (Fig.~\ref{fig:exp}(a)). A profile analysis of this pattern yields a periodicity of \SI{10}{\um}, consistent with that specified by piezoresponse force microscopy (PFM). Theoretically, the abrupt electric field change at a domain wall is expected to produce a sharp transition in the gradiometry. However, the absence of such distinct transitions in our results (Fig.~\ref{fig:exp}(b)) is likely attributed to surface screening\cite{kalinin_surface-screening_2018} or physisorption\cite{li_direct_2008}, which smooths the electric field profile from a sharp step into a gradual one. The surface screening on ferroelectric surfaces remain a challenge, even though the gradiometry technique can overcome the electrostatic screening originating from the diamond surface.

To quantify the spatial resolution of our scanning NV center electrometry, we perform repeated scans over a localized feature found at the center of a ferroelectric domain. The electric field map in Fig.~\ref{fig:exp}(d) reveals a stripe defect whose sharp edge provides an ideal test for the resolution. The defect is possibly originated from the pattern dynamics during domain switching\cite{ievlev_intermittency_2014}. We verify the feature's reproducibility and its electric-field origin. The feature is consistently observed with similar contrast and shape across multiple scans, confirming it is a stable, static feature rather than a transient artifact. Furthermore, a pulsed optically detected magnetic resonance (ODMR) measurement (Fig.~\ref{fig:exp}(e)) allows us to rule out magnetic artifacts. A high-resolution line scan across the steepest edge of the feature (Fig.~\ref{fig:exp}(f)) yields an experimental spatial resolution of approximately \SI{10}{\nm}, determined using the 10\%–90\% edge-width criterion. We simulate the point spread function (PSF) of a $\delta(x)$ charge distribution(Fig.~\ref{fig:exp}(g)). The PSF indicates that a \SI{10}{\nm} edge-width resolution occurs at an NV-sample distance of $\sim$\SI{17}{\nm}, where the corresponding resolution defined by the full width at half maximum (FWHM) is $\sim$\SI{15}{\nm} (Fig.~\ref{fig:exp}(h)).

Scanning NV center microcopy has been shown to be an appropriate scheme for electric field imaging applications. In this work, by implementing gradiometry in an intermittent-contact mode, we successfully suppress the mechanical oscillation amplitude of the tip from tens of nanometers to subnanometers, enabling the closer engagement of the NV center to the sample while keeping stable feedback of the distance. Finally, we achieve a spatial resolution of approximately \SI{10}{\nm} under ambient conditions. 

This tenfold improvement in spatial resolution, from $\sim$\SI{100}{\nm} down to $\sim$\SI{10}{\nm}, enables the detection of fine domain structures, such as those in Van der Waals ferroelectrics\cite{wang_interfacial_2022, weston_interfacial_2022}. Because these domains are typically tens of nanometers in size\cite{mcgilly_visualization_2020}, the enhanced resolution allows for the real-space mapping of stray electric fields from these previously unresolvable structures, thereby enabling the direct quantitative measurement of their local polarization. Furthermore, this capability would be conductive to the exploration of theoretically predicted topological polar structures\cite{chen_recent_2021}, such as meron-antimerons\cite{bennett_polar_2023}. This work provides a powerful tool to resolve the complexities of electric field distributions at the nanoscale, offering new opportunities for ferroelectric investigations.

\section{acknowledgments}
This work is supported by National Key R\&D Program of China (Grant Nos. 2023YFF0718400), the National Natural Science Foundation of China (Grants No. T2325023, 92265204, 12474500), the CAS (Grant Nos. ZDZBGCH2021002), Innovation Program for Quantum Science and Technology (Grant Nos. 2021ZD0303204, 2021ZD0302200) and USTC Tang Scholar.

\end{document}


\preprint{AIP/123-QED}

\title[Supplement]{Supplement}
\maketitle

\section{\label{sec:level1}Time synchronization}
\begin{figure}
\includegraphics{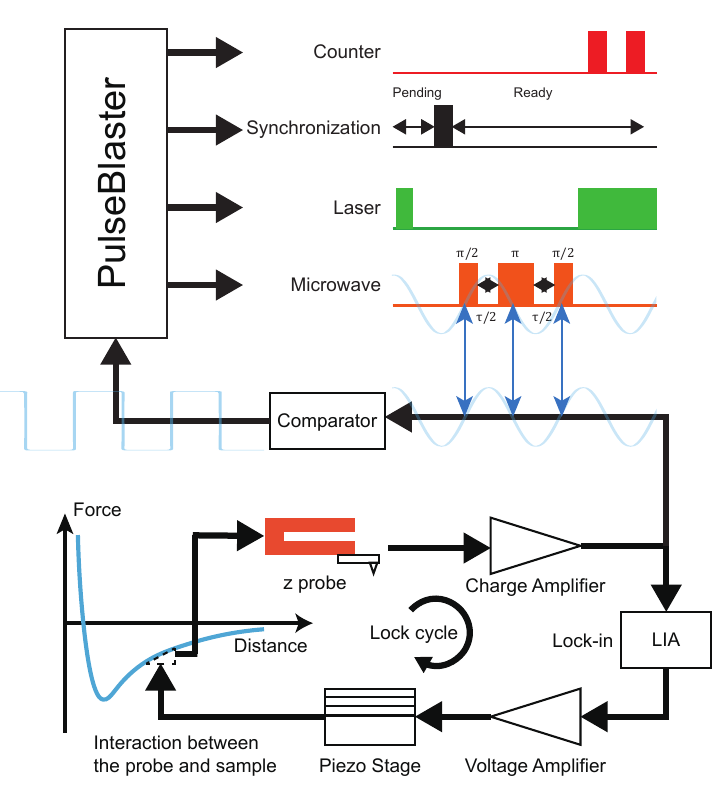}
\caption{\label{fig:time-synchronization} Time synchronization system for gradiometry.}
\end{figure}
In our setup, we employ a quartz tuning fork as the mechanical resonator. The fundamental resonance frequency appears at approximately 32 kHz, and the second-order resonance peak is around 180 kHz. In our experiments, we utilize the second-order resonance for force detection, where the resonance period is approximately 5.5 $\mu$s.
Gradiometry demands precise time-domain synchronization between the quantum control of NV center electronic spins in diamond and the mechanical oscillation of the probe. The $\sim$\SI{100}{\ns} quantum NOT gate must be applied precisely when the AC field changes the sign. 

Our synchronization system is illustrated in Fig.~\ref{fig:time-synchronization}.  The tuning fork for probe-sample distance feedback operates in the amplitude modulation mode (AM). The tuning fork is driven by a lock-in amplifier (LIA, Zurich Instruments MFLI) and detected after a home-made differential preamplifier. The PID result of the LIA drives the $z$-axis of a piezo stage (COREMORROW P63.XY18Z8S-D1) after a high-voltage amplifier (COREMORROW E70.D3S). The $x$- and $y$-axes in closed-loop configuration are used for lateral scanning, which is controlled by a multifunctional I/O device (National Instruments PCIe-6363). Following the charge amplifier, a splitter directs an additional route to a comparator, which converts the analog sine wave into a digital square wave. This square wave is adopted to trigger a pulse blaster (PB, Spincore PulseBlaster ESR-PRO-500), ensuring precise timing synchronization across all devices.

\section{\label{sec:level1}Simulation of Spatial Resolution}
We simulate the spatial resolution by considering a line charge, modeled as a surface charge distribution of $\delta(x)$. The resolution is determined by applying the 10\%-90\% edge-width criterion to the electric field profile across the charge.
We calculate the electric field distribution from the charge distribution in Fourier space:
\begin{equation}
\mathcal{F}\left(\bm{E}\right)=\frac{e^{-kz}}{4\pi\epsilon_0}\mathcal{F}\left(\sigma\right)\left(-\frac{ik_x}{k},-\frac{ik_y}{k},-1\right)
\end{equation}
where $\mathcal{F}$ denotes the Fourier transform, $k_x$,$k_y$ are the basis vectors in Fourier space, $k=\sqrt{k_x^2+k_y^2}$,$\epsilon_0$ is the vacuum permittivity, and $\sigma=\delta(x)$ is the surface charge distribution.
The Fourier transform of the charge distribution is:
\begin{equation}
    \mathcal{F}\left(\sigma\right) = \mathcal{F}\left[\delta(x)\right] = \delta(k_y)
\end{equation}
Substituting this into the previous equation for $\mathcal{F}\left(\bm{E}\right)$ and performing an inverse Fourier transform, we obtain the real-space components:
\begin{equation}
    E_{x} = - \frac{x}{4\pi^{2}\epsilon_{0}\left( {x^{2} + z^{2}} \right)}
\end{equation}
\begin{equation}
    E_{z} = - \frac{z}{4\pi^{2}\epsilon_{0}\left( {x^{2} + z^{2}} \right)}
\end{equation}
Furthermore, by symmetry:
\begin{equation}
    E_{y} = 0
\end{equation}

\begin{figure}
\includegraphics{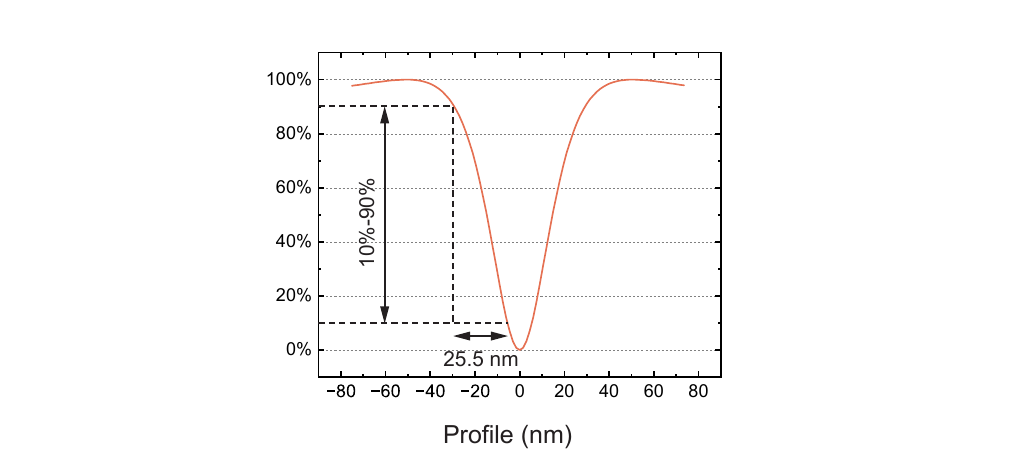}
\caption{\label{fig:intermittent-contact-resolution} Simulated scanning NV center profile for intermittent-contact mode.}
\end{figure}
We numerically simulate an NV center undergoing simple harmonic motion in the aforementioned electric field to emulate the experimental gradiometry. Simulating the mechanical oscillation, rather than using a gradient approximation, is crucial to avoid potential distortions at large amplitudes. For an NV-sample distance of \SI{30}{\nm} and a mechanical oscillation amplitude of \SI{10}{\nm}, the resulting simulated image is shown in Fig.~\ref{fig:intermittent-contact-resolution}. We define the resolution as the 10\%-90\% transition width of the edge profile. The relationship between the resolution, NV-sample distance, and mechanical amplitude is discussed in the main text.

\begin{figure}
\includegraphics{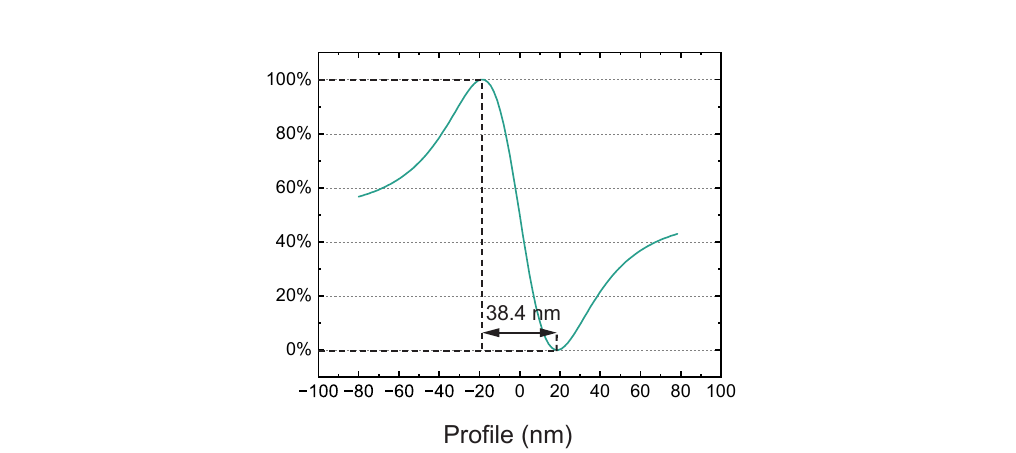}
\caption{\label{fig:shear-force-resolution} Simulated scanning NV center profile for shear-force mode.}
\end{figure}
Additionally, we simulate the gradiometry for the shear-force mode under the same electric field distribution. The result is shown in Fig.~\ref{fig:shear-force-resolution}, based on an NV-sample distance of \SI{30}{\nm} and a mechanical oscillation amplitude of \SI{10}{\nm}. The transverse oscillation is oriented along the x-direction; an oscillation along the y-direction would yield no gradiometry contrast, as the field is uniform along that axis. Notably, the resulting line profile for the shear-force mode does not exhibit a simple peak-like feature. Therefore, we define its resolution as the width of the sharpest transition region. The dependence of the resolution for the shear-force mode is discussed in the main text.

Note: We clarify that using this model, the resolutions of the intermittent-contact and shear-force modes cannot be directly compared due to the different definitions.

\section{\label{sec:level1}Design of the probe}
\begin{figure}
\includegraphics{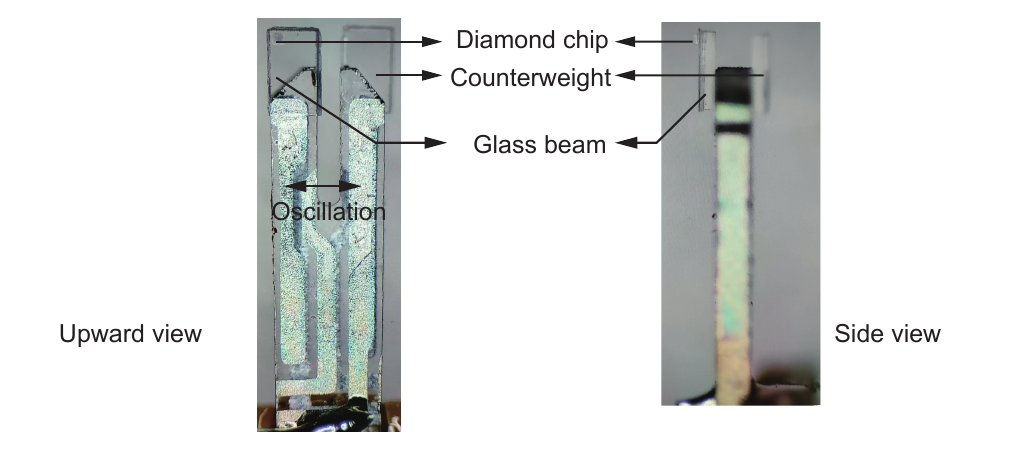}
\caption{\label{fig:shear-force-probe} Probe for shear-force mode.}
\end{figure}

\begin{figure}
\includegraphics{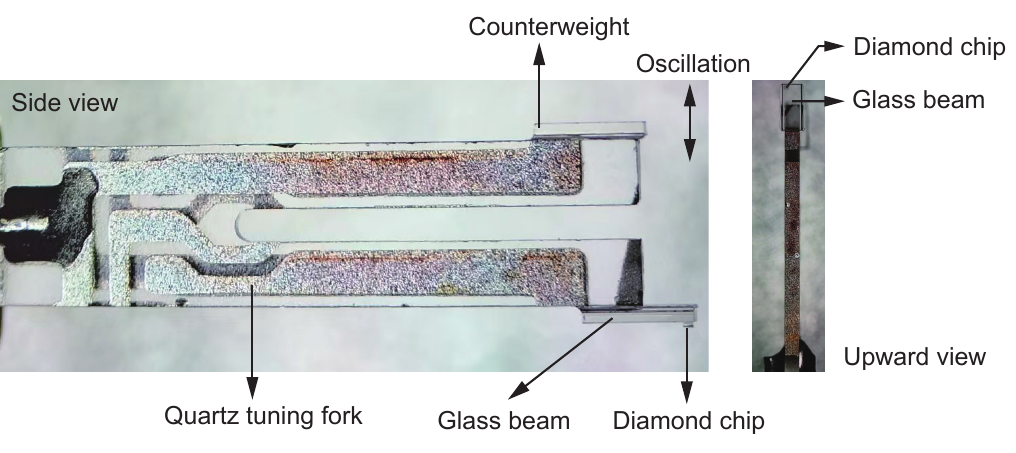}
\caption{\label{fig:intermittent-contact-probe} Probe for intermittent-contact mode.}
\end{figure}
At the second harmonic frequency, the tuning fork is prone to off-axis oscillations, even from a slight mass imbalance on its two prongs. This issue is particularly pronounced in the commonly used shear-force mode, where a diamond chip is attached to one of the prongs, making it difficult to trim the balance.

To adopt intermittent-contact mode, we attach a glass beam containing the diamond chip to the lower prong of a vertically oscillated quartz tuning fork. A counterweight is attached to the opposing upper prong, and its mass and position are carefully adjusted to maintain oscillation symmetry. This configuration not only improves the quality factor but also balances the oscillation at the second harmonic frequency, as we verify in the simulation.

The photos of our probes for shear-force mode and intermittent-contact mode are shown in Fig.~\ref{fig:shear-force-probe} and Fig.~\ref{fig:intermittent-contact-probe}, respectively.

\section{\label{sec:level1}Supplemental profile data}
\begin{figure}
\includegraphics{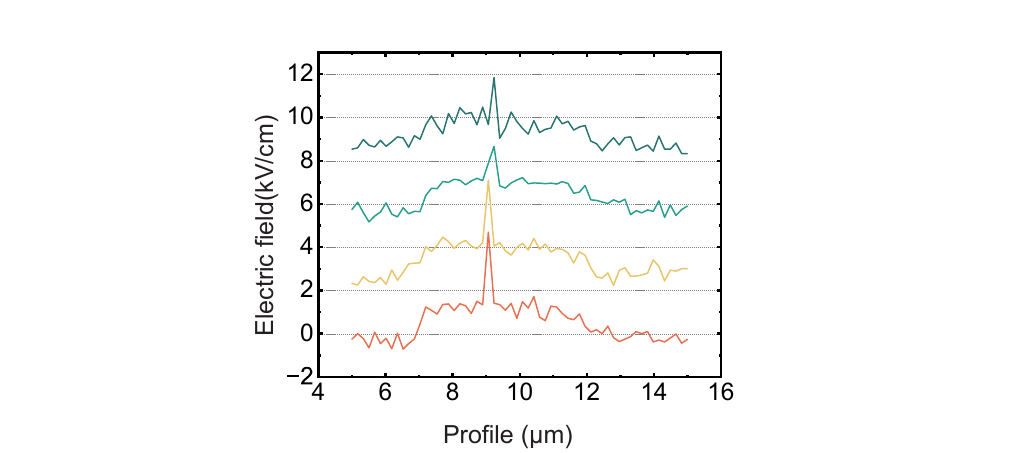}
\caption{\label{fig:large-data} Supplemental profile data for the defect from a \SI{10}{\um} scan.}
\end{figure}

\begin{figure}
\includegraphics{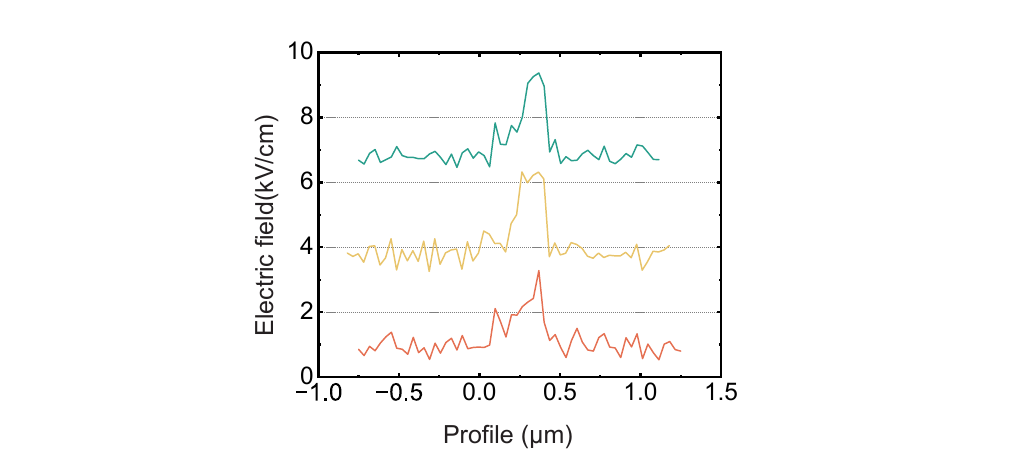}
\caption{\label{fig:middle-data} Supplemental profile data for the defect from a \SI{2}{\um} scan.}
\end{figure}

Supplemental profile data for the defect are shown in Fig.~\ref{fig:large-data} and Fig.~\ref{fig:middle-data}. The full fields of view for these profiles are \SI{10}{\um} and \SI{2}{\um}, respectively.

%